\documentstyle[12pt]{article}

\begin{document}

\vspace*{2cm}


\thispagestyle{empty}

\bigskip
\begin{center}
{\large \bf Mesons, Quarks and Leptons}\footnote[1]{To be published in a
commemorative volume of Bologna University in
honor of N. Zichichi}\footnote[2]{Supported by Deutsche Forschungsgemeinschaft,
DFG--Nr. FR 412/25--2}
\end{center}
\bigskip
\bigskip
\bigskip
\bigskip
\begin{center}

{\bf Harald Fritzsch}

\end{center}

\begin{center}

{\bf Sektion Physik, Ludwig--Maximilians--Universit\"at}\\

{\bf D--80333 M\"unchen, Germany}

\end{center}
\bigskip
\bigskip
\bigskip
\bigskip
{\bf Abstracts}
The QCD anomaly leads to an abnormal mixing and mass patter for the
pseudoscalar mesons. Furthermore it is responsible for the quality of
isospin symmetry in the meson spectrum. Similarities between the large
mixing angles among the neutral $0^{-+}$--mesons and the large mixing
angles observed in neutrino oscillations are pointed out.\\
\newpage
\noindent
In the field of strong interaction physics there has been a crucial
break\-through since the mid--sixties, the time when the first studies of
A. Zichichi and his collaborators on the radiative decays of mesons were
published. Let me describe shortly the situation at that time. A lot of
facts about the nature of the nuclear force were already known, e. g. the
strength of the interaction between nucleons, the magnitudes of the vertices
among nucleons and $\pi$-- or $K$--mesons or the radii and the
electromagnetic static properties as well as the form factors of the
nucleons.\\
\\
The success of $SU(3)$ symmetry, introduced in 1961 by Gell--Mann$^{1)}$ and
Neeman$^{2)}$, showed that in the physics of the strongly interacting
particles there
exist simple symmetry rules, although it remained mysterious why there
exists such a symmetry at all. What was even more mysterious was the fact that
the $SU(3)$ symmetry was broken at a level of about 20\%, but the symmetry
breaking showed a remarkable regularity -- it seemed to agree to a large
extent with the hypothesis that the $SU(3)$--violating term transformed like
an octet under the symmetry. This assumption leads to the Gell--Mann--Okubo
mass formula, which in case of the baryon octet and decuplet agree well with
experiment$^{3)}$.\\

In 1964 Gell--Mann$^{4)}$ in Pasadena and Zweig$^{5)}$ at CERN introduced the
quark model,
the hypothesis that all strongly interacting particles consist of three basic
spin 1/2 objects, called the $u, d$ and $s$--quarks. This model was able to
explain many observed features of the hadronic spectrum, but nevertheless was
received with great scepticism by the physics community, mostly due to the
fact that the quarks had to carry non--integral electric charges and no
mechanism was known which could explain why the baryons and mesons had the
structure ($qqq$) and $\left( \bar q q \right)$ respectively, while other
states like ($qq$) or $\left( \bar qqq \right)$ did not seem to exist. It
took eight years until a resolution of this paradox was found -- the theory
of QCD, which was based on the introduction of the new color degree of
freedom and its use as a gauge symmetry.\\

Zweig$^{5)}$ introduced the quarks (which he called aces at the time) by
writing down simple wave functions of the vector mesons in terms of
$\bar q q$--states:
\begin{eqnarray}
\rho & = & \frac{1}{\sqrt{2}} \left( \bar u u - \bar d d \right) \nonumber \\
\omega & = & \frac{1}{\sqrt{2}} \left( \bar u u + \bar d d \right) \nonumber \\
\Phi & = & \bar s s
\end{eqnarray}
In terms of quarks the $\Phi$--meson is a pure $\bar s s $--state. In terms
of the $SU(3)$--symmetry, however, it is a mixture of an $SU(3)$--octet and
an $SU(3)$--singlet:
\begin{eqnarray}
\Phi & = & {\rm cos} \, \Theta \cdot \frac{1}{\sqrt{6}} \left( \bar u u +
\bar d d - 2 \bar s s \right) + {\rm sin} \, \Theta \cdot \frac{1}{\sqrt{3}}
\left( \bar u u + \bar d d + \bar s s \right) \nonumber \\
\omega & = & - {\rm sin} \Theta \cdot \frac{1}{\sqrt{6}} \left( \bar u u +
\bar d d - 2 \bar s s \right) + {\rm cos} \, \Theta \cdot \frac{1}{\sqrt{3}}
\left( \bar u u + \bar d d + \bar s s \right)
\end{eqnarray}
where the mixing angle $\Theta $ is given by $\varphi = - {\rm arc \, tan}
\sqrt{\frac{1}{2}} = - 35.3^{\circ}$. Zweig observed that this angle is
remarkably close to the mixing angle obtained by using $SU(3)$ symmetry
for the vector meson octet and taking into account the Gell--Mann--Okubo mass
formula.
Thus the quark model provided a very simple picture for the neutral vector
mesons -- they segregate into the nearly degenerate $\rho_0 - \omega $ mesons
composed of $u$-- and $d$--quarks and
the $\Phi$--meson with the quark composition $\bar ss$. As found during the
sixties, especially by the Zichichi group, this simple picture of the vector
mesons agrees beautifully with the experimental data, in particular those
about radiative decays of mesons and about strong decays like
$\Phi \rightarrow \bar KK$.\\

The unusual structure of the vector mesons, given by their peculiar mixing
pattern when viewed from the $SU(3)$ platform, gives strong support that these
mesons consist of constituents, which in a strong interaction process like
$K^- + p \rightarrow \Lambda + \Phi \rightarrow \Lambda + K^- + K^+$ flow
smoothly between the vertices:
\begin{equation}
\left( \bar u s \right) + \left( u u d \right) \rightarrow \left( uds \right)
+ \left( \bar s s \right) \rightarrow \left( uds \right) + \left( \bar u s
\right) + \left( \bar s u \right) \, .
\end{equation}
In a 1999 recollection about one of the most important discoveries of the
20th century George Zweig writes$^{6)}$:\\
``In the April 15, 1963 Physical Review Letters there is a paper$^{7)}$
ti\-tled
``Existence and Properties of the $\Phi$--Meson''. I remember being very
surprised by Figure 1, which showed a Dalitz plot for the reaction $K^- + p
\rightarrow \Lambda + K + \bar K$. There was an enormous peak at about
$\left( 1020 MeV\right)^2$ in the $M^2$--plot for $K \bar K$, right at the
edge of phase
space. The fact that the $\Phi$ decayed predominately into $K \bar K$ and not
$\rho \pi$ was totally unintelligible despite the authors' assurance that this
suppression ``need not be disconcerting''. A spin one $\Phi$ would decay into
$\rho \pi$ or $K \bar K$ in a $P$--wave. Since the $\Phi$ was just
slightly above the $K \bar K$--threshold, the $P$--wave $K \bar K$--mode was
highly suppressed. My estimate indicated that the $\rho \pi$--decay mode was
at least two orders of magnitude below what might be expected.
Feynman taught me that in strong interaction physics everything that possibly
can happen does, and with maximum strength. Only conserva\-tion laws suppress
reactions. Here was a reaction that was allowed but did not proceed! I had
thought that hadrons probably have constituents, and this experiment
convinced me that they do, and that they are real. By assigning the proper
constituents to the pseudoscalar and vector mesons, and assuming that when
hadrons decay their constituents flow into the particles they decay into, the
striking $\rho \pi$ suppression in $\Phi$ decay could be ``explained''. This
was a statement about dynamics which indicated that the constituents were not
hypothetical objects carrying the symmetries of the theory, but real objects
that moved in space--time from hadron to hadron! Later when I explained this
to Feynman at Nino's 1964 Erice summer school$^{8)}$ where we both were
speakers, Feynman did not believe my argument because ``unitarity connects all
channels with the same quantum numbers, so $\rho \pi$ and $K \bar K$ get all
mixed up, making the suppression of $\rho \pi$ impossible''. Feynman
thought the
experiment was wrong, or that there was something else going on that we didn't
understand. Later that fall, when I gave Gell--Mann my explanation of
$\Phi $ decay and drew my diagram for $\Phi \rightarrow \rho + \pi$ (which
involved polygonal block--like icons for the constituents), I can still hear
him saying ``Oh, the concrete quark model!''.''\\

Nevertheless the naive quark model is confronted with a severe problem. It
does not explain why the neutral vector mesons segregate in the way they do.
In principle the strong interaction can easily cause transitions between the
neutral
$\left( \bar qq \right)$--systems, and they could be such that the mixing
angle between $\omega $ and $\Phi$ deviates considerably from the ``naive''
angle discussed above. Apparently such transition
amplitudes are nearly absent for the neutral vector mesons. This is part of
the mystery surrounding the dynamics of the Zweig rule until today.\\

In QCD violations of the Zweig rule, e. g. the decay $\Phi \rightarrow 3 \pi$,
proceed through gluonic intermediate states, carrying the quantum numbers
$J^{PC} = 1^{--}$. In lowest order of perturbation theory a three--gluon
configuration is the simplest color--singlet gluonic configuration which can
contribute. Formally the transition amplitude between a $\left( \bar u u
\right)$
and $\left( \bar s s \right)$--state is of order $\left( \alpha _s \right)^3$,
where $\alpha_s$ is the QCD coupling constant. If $\alpha _s$ is small
already at energies of the order of 1 GeV and if perturbative approximations
make sense at all, the Zweig rule could be interpreted this way. The question
arises, however, whether the perturbative arguments can be applied in
an energy region where nonperturbative phenomena and confinement effects are
certainly not negligible. In this sense even today the Zweig rule and the
mixing pattern of the neutral vector mesons is not completely understood.\\

In QCD one expects that the mass term of the neutral vector mesons is given
by$^{8)}$
\begin{equation}
M \left( \bar q q \right) = \left( \begin{array}{ccc}
                                M \left( \bar u u \right) & 0 & 0 \\
                                0 & M \left( \bar d d \right) & 0 \\
                                0 & 0 & M \left( \bar s s \right)
                                  \end{array} \right)
     + \sigma \left( \begin{array}{ccc}
                1 & 1 & 1\\
                1 & 1 & 1\\
                1 & 1 & 1
               \end{array} \right)
\end{equation}
where $\sigma $ is a parameter describing the strength of the mixing between
$\left( \bar u u \right)$, $\left( \bar d d \right) $ and
$\left( \bar s s \right)$ caused by the gluons. If $\sigma$ would be exactly
zero, the three mass eigenstates would have the quantum number of pure
$\left( \bar u u \right), \left( \bar d d \right)$ and $\left(\bar s s
\right)$--states. Note that the $\left( \bar u u \right)$ and
$\left( \bar d d \right)$--states are not exactly degenerate in mass, due to
the isospin violating $m_u - m_d$ mass difference.\\

We should like to note that the matrix multiplying $\sigma $ is taken to be
$SU(3)$ symmetric. In reality there are small $SU(3)$ violations, which will
not be discussed here.\\

If $\sigma $ would be large compared to $M \left( \bar u u \right)$ or
$M \left( \bar s s \right)$, the mass eigenstates would \, be close to the
states $\left( \bar u u - \bar d d \right) / \sqrt{2}$ and
$\left( \bar u u + \bar d d - 2 \bar s s \right) / \sqrt{6}$,\\i. e. the
isospin triplet and the $SU(3)$ octet state. In reality we have an isospin
triplet, the $\rho\,  (770)$, the $\omega $ (783) which is nearly
$\left( \bar u u + \bar d d \right) / \sqrt{2}$, and a state which is nearly
$\left( \bar s s \right)$, the $\Phi$(1020). Such a pattern can only be
achieved, if the mixing term $\sigma $ is very small compared to the mass
difference $M \left( \bar ss \right) - M \left( \bar u u \right)$. If
$\sigma $ were zero, the $\Phi$--meson would be a pure
$\left( \bar ss \right)$--state, which is not the case. The radiative decay
$\Phi \rightarrow \pi^0 \gamma $, observed with a branching ratio of about
0.0013, can be used to estimate the amount of nonstrange quarks in the
$\Phi$--meson wave function. One finds:
\begin{equation}
\Phi \cong \bar s s + 0.06 \left( \bar u u + \bar d d \right) / \sqrt{2} \, .
\end{equation}
This gives an estimate for $\sigma $, which is of the order of 10 MeV. The
$\omega - \rho_0$ mass difference will also be given by $\sigma $. One expects
\begin{equation}
M ( \omega ) - M \left( \rho^0 \right) = 2 \sigma \approx 20 MeV.
\end{equation}
In reality this mass difference is about 12 MeV, not a bad estimate in
view of the fact that we have disregarded the effects of isospin violations.\\

Any difference between the diagonal terms $M \left( \bar u u \right)$ and
$M \left( \bar d d \right)$, which is a measure of isospin violation, would
cause a mixing between $\rho $ and $\omega$, i. e. the $\omega$--meson would
have a small admixture of the $\Delta I = 1$ term $\left( \bar u u - \bar d
d \right) / \sqrt{2}$. Using the decay $\omega \rightarrow \pi^+ \, \pi^-$,
observed with branching ratio of 2.2 \%, one estimates this admixture to be
about 3.5 \% in amplitude, which implies that the mass difference
$M \left( \bar u u \right) - M \left( \bar d d \right)$ should be less than
1 MeV.\\

As soon as $m_s - m_{u, d}$ is lifted from zero, a singlet--octet mixing
sets in. An interesting situation would arise if $M \left( \bar s s \right) -
M \left( \bar u u \right)$ is equal to $\sigma $. In this case the two states
$\omega $ and $\Phi$ are given by a 45$^{\circ}$ rotation:
\begin{eqnarray}
\omega & = & \frac{1}{\sqrt{2}} \left[ \left( \left( \bar u u + \bar d d
\right) / \sqrt{2} - \bar s s \right) \right] \nonumber \\
\Phi & = & \frac{1}{\sqrt{2}} \left[ \left( \bar u u + \bar d d / \sqrt{2}
\right) + \bar s s \right] \, .
\end{eqnarray}
Of course, this is a hypothetical case, since in reality $\sigma $ is much
smaller than $M \left( \bar s s \right) - M \left( \bar u u \right)$. In
reality we have:
\begin{eqnarray}
\omega & \approx & \left( \bar u u + \bar d d \right) / \sqrt{2} - 0.06 \cdot
\bar s s \nonumber \\
\Phi & \approx & 0.06 \left( \bar u u + \bar d d \right) / \sqrt{2} +
\bar s s \, .
\end{eqnarray}

Furthermore it is useful to consider the $SU(3)$--limit $m_u = m_d = m_s$
in the
absence of electromagnetism. Due to
the $SU(3)$--symmetry the neutral vector mesons would segregate into the two
mesons $ \left( \bar u u - \bar d d \right) / \sqrt{2}$ and $\left( \bar u u
+ \bar d d - 2 \bar s s \right) / \sqrt{6}$, degenerate in mass, and the
$SU(3)$ singlet $\left( \bar u u + \bar d d + \bar s s \right) / \sqrt{3}$,
which is lifted in its mass due to the mixing term. The resulting mass
difference is given by 3 $\sigma$, about 30 MeV.\\

While the mixing strength caused by the gluonic interaction is small, but
non--zero in the $J^{--}$--channel, a different situation arises in the
$0^{-+}$--channel, i. e. for neutral pseudoscalar mesons. In the limit
$m_u = m_d = m_s$ QCD exhibits a chiral $SU(3) \times SU(3)$ symmetry. As a
result
there are eight massless pseudoscalar Goldstone bosons in the symmetry limit,
the three pions, four kaons and the $\eta$--meson. The ninth meson, the
$\eta'$--state, acquires a mass due to the gluonic interaction.
Phenomenologically the (mass)$^2$--matrix of the neutral pseudoscalar mesons
can be written as$^{9)}$
\begin{equation}
M^2 \left( \bar q q \right) = \left( \begin{array}{ccc}
                                M^2 \left( \bar u u \right) & 0 & 0 \\
                                0 & M^2 \left( \bar d d \right) & 0 \\

0& 0 & M^2\left(\bar ss \right)
                                \end{array} \right)
+ \lambda \left( \begin{array}{ccc}
                  1 & 1 & 1\\
                  1 & 1 & 1\\
                  1 & 1 & 1
                \end{array} \right)~~~.
\end{equation}

Here the mixing strength is described by the parameter $\lambda $. The mass
terms $M^2 \left( \bar u u \right)$ etc. are proportional to $m_u$ etc.,
according to the chiral symmetry constraints.\\

In the limit of chiral symmetry $m_u = m_d = m_s = 0$ the mass squared of the
singlet
state $\eta '$ (quark composition $\left( \bar u u + \bar d d + \bar s s
\right) / \sqrt{3}$) is given by $ 3 \lambda$. In the limit of isospin
invariance $\left( m_u = m_d \right)$ we take
\begin{eqnarray}
M^2 \left( \bar u u \right) & = &
M^2 \left( \bar d d \right) = M^2_{\pi } \cong 0.02~{\rm GeV}^2, \nonumber \\
M^2 \left( \bar s s \right) & = & 2 M^2 \left( \bar s u \right) - 2 M^2
\left( \bar u u \right) = 2 M^2_K - 2 M^2_{\pi} \cong 0.45 ~{\rm GeV}^2
\end{eqnarray}
and we obtain:
\begin{equation}
\lambda = \frac{1}{3} \left( M^2_{\eta } + M^2_{\eta '} - M^2
\left( \bar s s \right) - M^2 \left( \bar u u \right) \right) \cong ~{\rm
GeV}^2
\end{equation}

\bigskip

One finds $M_{\eta} \cong 500 $ MeV, $M_{\eta '} \cong 980$ MeV, in a
reasonable agreement with the observed values
\begin{equation}
M_{\eta } \cong 547 \, MeV, \, M_{\eta '} \cong 958 \, \, MeV \, .
\end{equation}
Note that we have assumed that the gluonic mixing term is $SU(3)$ invariant.
However $SU(3)$ breaking effects will also affect the gluonic transitions,
i. e. the transition terms $\left( \bar u u \leftrightarrow \bar d d \right)$
and $\left( \bar u u \leftrightarrow \bar s s \right)$ will differ slightly.
These $SU(3)$ breaking effects will also change the mass eigenvalues
slightly.
Neglecting these effects, one finds the wave functions:
\begin{eqnarray}
\eta & \cong & 0.79 \left( \bar u u + \bar d d \right) / \sqrt{2} - 0.62
\bar s s \nonumber \\
\eta' & \cong & 0.62 \left( \bar u u + \bar d d \right) / \sqrt{2} + 0.79
\bar s s \, .
\end{eqnarray}
Thus a rather strong mixing between the various
$\left( \bar q q \right)$--configurations is observed, which is a consequence
of the gluonic anomaly of QCD$^{10, 11)}$ and on the phenomenological side a
consequence \,
of the large \, mixing term given by $\lambda $.\\

The actual value of the mixing angle in the $0^{-+}$--sector is not yet
known with good precision. For many years a singlet--octet mixing angle of
$10^{\circ}$ was assumed. This gives:
\vfill\eject
\begin{displaymath}
| \eta > = {\rm cos} \, 10^{\circ} \left( \bar u u + \bar d d - 2 \bar s s
\right) / \sqrt{6} + {\rm sin} \, 10^{ \circ} \left( \bar u u + \bar d d +
\bar s s \right) / \sqrt{3}
\end{displaymath}
\begin{displaymath}
\cong 0.71 \left( \bar u u + \bar d d \right) / \sqrt{2} - 0.70
\bar ss
\end{displaymath}
\begin{equation}
| \eta' > = 0.70 \left( \bar u u + \bar d d \right) / \sqrt{2} + 0.71 \bar ss
\, .
\end{equation}
These wave functions are close to the wave functions given by Feynman$^{12)}$:
\begin{eqnarray}
| \eta > & = & \frac{1}{2} \left( \bar u u + \bar d d - \sqrt{2} \bar s s
\right) \nonumber \\
| \eta ' > & = & \frac{1}{2} \left( \bar u u + \bar d d + \sqrt{2} \bar s s
\right) \, .
\end{eqnarray}
These wave functions are such that the $\bar s s$--term changes sign in the
transition from $\eta $ to $\eta '$. They correspond to a
45$^{\circ}$--rotation between
$\bar s s $ and $\left( \bar u u + \bar d d \right) / \sqrt{2}$, the
hypothetical case discussed above for the vector mesons (eq. (7)). Another
interesting set of wave functions is:
\noindent
\begin{eqnarray}
| \eta > & = & \frac{1}{\sqrt{3}} \left( \bar u u + \bar d d - \bar s s
\right) \nonumber \\
| \eta' > & = & \frac{1}{\sqrt{6}} \left( \bar u u + \bar d d + 2 \bar s s
\right) \, ,
\end{eqnarray}
which corresponds to a singlet--octet mixing angle of 19.5$^{\circ}$. Here
the $\eta \left( \eta ' \right)$--wave functions are obtained from the
singlet (octet) states by just switching the sign of the $\bar ss $--term.
In reality the mixing angle seems to be between these two cases: $\Theta
\approx 15^{\circ}$, and we have:
\begin{eqnarray}
| \eta > & = & 0.77 \left( \bar u u + \bar d d \right) / \sqrt{2} - 0.64
\left( \bar ss \right) \nonumber \\
| \eta ' > & = & 0.64 \left( \bar u u + \bar d d \right) / \sqrt{2} + 0.77
\left( \bar s s \right) \, .
\end{eqnarray}
The question remains open whether the wave functions of the $\eta '$--meson
(and to a lesser extent also of the $\eta$--meson) have small admixtures of
gluonic configurations. But in any case the strong mixing among
the neutral $0^{-+} - \bar qq$--configurations indicates that unlike the
vector meson channel the pseudoscalar channel is particularly sensitive to
the dynamics of QCD. The mixing strength in the $0^{-+}$--channel is much
stronger than the mixing strength in the $1^{--}$--channel such that a
qualitatively new situation arises.

In the chiral limit $m_u = m_d = m_s = 0$ the mass spectrum of the neutral
pseudoscalars exhibits two zero eigenvalues and a non--zero one
$\left( M^2_{\eta '} = 3 \lambda \right)$. Thus a strong mass hierarchy
exists. The mass matrix in the basis given by the $\pi^0, \eta$ and
$\eta'$--states is given by:
\begin{equation}
M = {\rm const.} \, \, \, \, \left( \begin{array}{ccc}
                                0 & 0 & 0\\
                                0 & 0 & 0\\
                                0 & 0 & 1
                                  \end{array} \right) \, .
\end{equation}
This mass matrix is of rank one and can be rewritten as follows:
\begin{equation}
M = {\rm const.} \, \, \, \,  \left( \begin{array}{ccc}
                                1 & 1 & 1\\
                                1 & 1 & 1\\
                                1 & 1 & 1
                                   \end{array} \right) \, ,
\end{equation}
using a $3 \times 3$ matrix, whose elements are universal  -- a
consequence of the universality of the gluonic force in QCD. The mass
matrix takes this form if one performs a unitary transformation
among $
\pi^0,
\eta$ and $\eta '$. According to QCD the new states are the $\bar q q
$--bilinears such that:
\begin{eqnarray}
\pi^0 & = & \left( \bar u u - \bar d d \right) / \sqrt{2} \nonumber \\
\eta  & = & \left( \bar u u + \bar d d - 2 \bar s s \right) / \sqrt{6}
\nonumber \\
\eta' & = & \left( \bar u u + \bar d d + \bar s s \right) / \sqrt{3} \, .
\end{eqnarray}
If one uses the $\left( \bar qq \right)$--states instead of the mass
eigenstates, the mass matrix exhibits an $S (3)$--symmetry, which can only
be seen in that basis. The mass matrix is invariant under a flavor
permutation, e.g., under the permutation $(\bar uu)\leftrightarrow (\bar
dd)$. In the case of three flavors the symmetry is $S(3)$, the discrete
symmetry of permutations of three elements. (In our case these are the
flavor eigenstates $(\bar uu)$, $(\bar dd)$ and $(\bar ss)$.) Since the
mass matrix in terms of the flavor eigenstates is proportional to a
$3\times 3$ matrix, in which all elements are equal, we denote the
$S(3)$-symmetry as the symmetry of flavor universality.
Thus the states $\pi^0, \eta, \eta'$ are mass eigenstates, while
the states
$\left( \bar u u \right), \left( \bar d d \right), \left( \bar s s \right)$
are the eigenstates of the flavor universality, but not mass
eigenstates.

The observed mass spectrum of the pseudoscalar mesons can
be seen in analogy to the mass spectra of the charged leptons and quarks.
The mass spectra of the charged leptons and quarks are dominated
essentially by the masses of the
members of the third family, i.\ e.\ by $\tau$, $t$ and $b$. Thus a clear
hierarchical
pattern exists. Furthermore the masses of the first family are small
compared
to those of the second one. Moreover, the CKM--mixing matrix exhibits a
hierarchical pattern -- the transitions between the second and third
family
as well as between the first and the third family are small compared to
those between the first and the second family.\\

The observed hierarchies signify that nature seems to be close to the
so--called ``rank--one'' limit, in which all mixing angles vanish and both
the $u$-- and $d$--type mass matrices and the  charged leptons are proportional
to the hierarchical rank-one matrix $M_0^h$:
\begin{equation}
M^h_0 = {\rm const.} \cdot \left(\begin{array}{ccc}
0 & 0 & 0\\ 0 & 0 & 0\\ 0 & 0 & 1 \end{array} \right) \, .
\end{equation}

\hspace*{0.1cm}
Whether the dynamics of the mass generation allows that this limit can
be
achieved in a consistent way remains an unsolved issue, depending on the
dynamical details of mass generation. Encouraged by
the
observed hierarchical pattern of the masses and the mixing parameters,
we
shall assume that this is the case. In itself it is a non-trivial
constraint
and can be derived from imposing a chiral symmetry for the massless
flavors$^{13)}$.
This symmetry ensures that an electroweak doublet which is massless
remains
unmixed and is coupled to the $W$--boson with full strength. As soon as
the mass is introduced, at least for one member of the doublet, the symmetry
is violated and mixing phenomena are expected to show up. That way a chiral
evolution of the CKM matrix can be constructed. At the first stage
only the $t$ and $b$ quark masses are introduced, due to their
non-vanishing
coupling to the scalar ``Higgs'' field. The CKM--matrix is unity in this
limit. At the next stage the second generation acquires a mass.
Since
the $(u, d)$--doublet is still massless, only the second and the third
 generations
mix, and the CKM--matrix is given by a real $2 \times 2$ rotation matrix
in the
$(c, \, s) - (t, \, b)$ subsystem, describing the flavor transitions between
the second and third family. Only at the next step, at which the $u$ and $d$
masses are introduced, does the full CKM--matrix appear, described in general
by three angles and one phase and only at this step $CP$--violation can
appear. Thus it is the generation of mass for the first family which is
responsible for the violation of $CP$--symmetry in the Standard Model.\\

The rank-one mass matrix can be expressed in terms of a matrix
exhibiting the flavor universality:
\begin{equation}
M_0 = c \left( \begin{array}{ccc}
1 & 1 & 1\\ 1 & 1 & 1\\ 1 & 1 & 1 \end{array} \right) \, ,
\end{equation}
Its symmetry is a  $S(3)_L \, \, \times \, \, S(3)_R$ symmetry. It is
obtained from $M_0^h$ by an orthogonal transformation. Writing down the
mass eigenstates in terms of the eigenstates of flavor universality, one
finds e.g. for the lepton channel:
\begin{eqnarray}
e^0 & = & \frac{1}{\sqrt{2}} (l_1 - l_2) \nonumber\\
\mu^0 & = & \frac{1}{\sqrt{6}} (l_1 + l_2 - 2l_3)\\
\tau^0 & = & \frac{1}{\sqrt{3}} (l_1 + l_2 + l_3) \nonumber .
\end{eqnarray}

Here $l_1, \ldots$ are the symmetry eigenstates. Note that $e^0$ and $\mu^0$
are mass\-less in the limit considered here, and any linear combination of the
first two state vectors given in eq. (3) would fulfill the same purpose,
i.\ e.\ the decomposition is not unique, only the wave function of the
coherent state $\tau^0$ is uniquely defined. This ambiguity will disappear as
soon as the symmetry is violated.

The $\lambda$--term in the mass matrix (9) describes the result of the
QCD--anomaly which causes strong
transitions between the quark eigenstates (due to gluonic annihilation
effects
enhanced by topological effects). Likewise one may argue that analogous
transitions are the reason for
the lepton--quark mass hierarchy. Here we shall not speculate about a
detailed mechanism of this type, but merely study the effect of symmetry and
symmetry breaking.\\

Just like for the leptons above universal mass matrices can also be
introduced both for the up-- and down--quarks. These mass matrices are
supposed to be valid in the limit where the first and second family of
leptons and quarks are massless. Small violations of the symmetry can
account for the masses of the second and first family of quarks as well
as for the flavor mixing angles. In a similar way one can introduce small
violations of flavor universality to account for the mass of the muon
and of the electron.\\

The question arises whether a similar
structure can be also imposed in the neutrino sector. In the symmetry
limit
 only the $\tau$--lepton acquires a mass. Suppose a mass would also be
introduced for the $\tau$--neutrino$^{14)}$ along the same line. In this
case we
would obtain a massive $\tau$--neutrino, which could also be a Majorana or a
Fermi--Dirac state, and the neutrino mass matrix takes the form:
\begin{equation}
M_{0\nu} \; = \; c_{\nu}  \left (
\begin{array}{ccc}
1       &       1       &       1 \\
1       &       1       &       1 \\
1       &       1       &       1 \\
\end{array} \right ) \; .
\end{equation}

Note that $m_{\nu_{\tau}} = 3c_{\nu}$. Since according to astrophysical
constraints the $\nu_{\tau }$--state must be rather light,
i.\ e.\ not heavier than about 30 eV, we would have a situation
in which the constants $c_{\nu } / c_{l}$ for the various flavor channels
differ by at least eight orders of magnitude
($c_{\nu } / c_{l}$ $ < 30 eV / m(\tau ) \approx 10^{-8})$. We find
such a tiny ratio rather unnatural, and one is invited to look
for other ways to introduce the neutrino masses.\\

In our view the simplest way to avoid the problem mentioned above is
to \, suppose that the constant \, $c_{\nu }$ \, vanishes, i.\ e.\ the \,
neutrinos do not
receive any mass contribution in the symmetry limit. One may
speculate about the dynamical reason for the vanishing of  $c_\nu$. For
example it would follow if one could establish a
multiplicative relation between the fermion masses in the symmetry
limit and their electric charges, i.e. the vanishing of $c_{\nu}$ would
be directly related to the fact that the neutrinos are electrically
neutral. If $c_{\nu}$ vanishes, it is automatically implied that
there exists a qualitative difference between the neutrino sector and the
charged lepton sector. In particular it may be not surprising that the
neutrino masses are small compared to the masses of the charged leptons. Thus
there would be no reason why the hierarchical pattern observed for the charged
lepton masses should repeat itself for the neutrino masses. The neutrino
masses could even be of the same order of magnitude.\\

In the absence of the universal neutrino mass term one would have:
\begin{eqnarray}
M_l & \; = \; & c_l \left (
\begin{array}{ccc}
1       &       1       &       1 \\
1       &       1       &       1 \\
1       &       1       &       1 \\
\end{array} \right ) + \Delta M_l \; ,
\nonumber \\
\nonumber \\
M_{\nu } & \; = \; & 0 + \Delta M_{\nu}
\end{eqnarray}
where $\Delta M_l$ and $\Delta M_{\nu}$ are the symmetry breaking terms for
the charged leptons and neutrinos, respectively. In the case of the
pseudoscalar mesons the breaking of the $S(3)$-symmetry, which
breaks also the chiral $SU(3) \times SU(3)$--symmetry, is given by diagonal
terms, the terms proportional to the quark masses. In a similar way the
simplest breaking term in the case of the leptons would be a diagonal mass
shift for the symmetry eigenstates, i. e.
\begin{eqnarray}
\Delta M_l & \; = \; & \left (
\begin{array}{ccc}
\delta_l       &       0       &       0 \\
0       &      \rho_l       &       0 \\
0       &       0       &       \varepsilon_l \\
\end{array} \right ) \; ,
\nonumber \\
M_{\nu} & \; = \; & \left (
\begin{array}{ccc}
\delta_{\nu}       &       0       &       0 \\
0       &      \rho_{\nu}       &       0 \\
0       &       0       &       \varepsilon_{\nu} \\
\end{array} \right ) \; .
\end{eqnarray}

Here both $\Delta M_l$ and $M_{\nu}$ are real matrices, i.\ e.\
$CP$--symmetry is
preserved for the leptons. Note that the neutrino mass matrix is already
diagonal ($\delta _{\nu }, \rho_{\nu}, \varepsilon _{\nu})$, while the mass
matrix for the charged leptons needs to be
diagonalized. Apart from small corrections, the main effect
of the diagonalization is to diagonalize the  $M_{0l}$
by the transformation $U M_{0l} U^{\dagger} = M^l_{H}$, where
$M^l_{H}$ is the ``hierarchical'' matrix:
\begin{equation}
M^l_H \; =\; c_l \left (
\begin{array}{ccc}
0  &  0  &  0      \\
0  &  0  &  0      \\
0  &  0  &  3
\end{array} \right ) \; ,
\end{equation}

and

\begin{equation}
U \; = \; \left (
\begin{array}{ccc}
\displaystyle\frac{1}{\sqrt{2}} & -\displaystyle\frac{1}{\sqrt{2}}      & 0 \\
\displaystyle\frac{1}{\sqrt{6}} & \displaystyle\frac{1}{\sqrt{6}}       &
-\displaystyle\frac{2}{\sqrt{6}} \\
\displaystyle\frac{1}{\sqrt{3}} & \displaystyle\frac{1}{\sqrt{3}}       &
\displaystyle\frac{1}{\sqrt{3}}
\end{array} \right ) \; ,
\end{equation}

\noindent
Thus the leptonic flavor mixing is essentially given by the rotation
matrix $U$ above, i.\ e.\ the leptonic
doublets are given by

\begin{equation}
\left (
\begin{array}{ccc}
\displaystyle\frac{1}{\sqrt{2}} (\nu_1 - \nu_2)
\quad & \quad \frac{1}{\sqrt{6}} \left( \nu_1 + \nu_2 - 2 \nu_3 \right)
\quad & \quad \displaystyle\frac{1}{\sqrt{3}} \left( \nu_1 + \nu_2 + \nu_3
\right) \\
e^- \quad &  \quad \mu^- \: & \:  \tau^-
\end{array} \right )
\end{equation}
\\

where $\nu_1, \nu_2, \nu_3$ are the neutrinos mass eigenstates.\\

We should like to mention another possibility to arrive at the mass pattern
given above. In the Standard Model neutrinos are massless Weyl fermions. The
only masses they could acquire are Majorana masses. In the limit of
flavor universality the simplest Majorana mass term is proportional to the
unit matrix:

\begin{equation}
M ( \nu) = {\rm const.} \, \, \left( \begin{array}{ccc}
                                   1 & & \\
                                   & 1 & \\
                                   & & 1
                                        \end{array} \right)
\end{equation}

This can be seen as follows. If neutrino were Fermi--Dirac objects, like the
charged fermions, their symmetry in the limit of flavor universality would be
$S(3)_L \times S(3)_R$. Left-handed and right-handed states could be
transformed
independently. However, for a Majorana fermion the left-handed and right-handed
states are linked by a $CP$--reflection since a four--component Majorana
fermion can be viewed as a combination of a left-handed fermion and a
right-handed antifermion.
Thus in the limit of $CP$--invariance the symmetry group is
not $S(3)_L \times S(3)_R$, but rather the diagonal sum $S(3)$.
Correspondingly the simplest mass matrix respecting the symmetry is not the
 matrix consisting of nine unit elements, but the diagonal
submatrix, i. e. the unit matrix.

Indeed a mass term given the same Majorana
mass to all three neutrinos is symmetric under $S(3)$. That way we arrive at
a structure similar to the one discussed above. In the symmetry limit the
mass matrices of the leptons are given by:
\begin{eqnarray}
M(l^-) & = & c_l \left( \begin{array}{ccc}
                        1 & 1 & 1\\
                        1 & 1 & 1\\
                        1 & 1 & 1
                        \end{array} \right)
\nonumber \\
\nonumber \\
M(\nu) & = & c_{\nu} \left( \begin{array}{ccc}
                        1 & 0 & 0\\
                        0 & 1 & 0\\
                        0 & 0 & 1
                        \end{array} \right) \, .
\end{eqnarray}
The symmetry violations would introduce small departures from this structure,
and we are back to the pattern discussed previously.\\

It is instructive to compare this situation with the pseudoscalar and
vector mesons. In the limit of chiral symmetry
$\left( m_u = m_d = m_s = 0 \right)$ the mass terms of the vector mesons and
pseudoscalar mesons exhibit also an $S(3)$--symmetry:
\begin{eqnarray}
M^2 \left( 1 ^{--} \right) & = & c \left( 1^{--} \right)
\left( \begin{array}{ccc}
        1 & 0 & 0 \\
        0 & 1 & 0 \\
        0 & 0 & 1
        \end{array} \right)
\nonumber \\
M^2 \left( 0 ^{-+} \right) & = & c \left( 0^{-+} \right)
\left( \begin{array}{ccc}
        1 & 1 & 1\\
        1 & 1 & 1\\
        1 & 1 & 1
        \end{array} \right) \, .
\end{eqnarray}

Here we have neglected the tiny gluonic mixing term in the $1^{--}$--channel.
This qualitative difference between the two mass terms introduces a
qualitative difference in the mixing pattern. The mass eigenstates of the
vector mesons are $\bar u u, \bar d d$ and $\bar s s$, the mass eigenstates of
the pseudoscalars are mixtures:
\begin{equation}
\left( \bar u u - \bar d d \right) / \sqrt{2}, \qquad \left( \bar u u + \bar
d d - 2 \bar s s \right) / \sqrt{2}, \qquad \left( \bar u u + \bar d d + \bar
s s \right) / \sqrt{3} \, .
\end{equation}

At the same time the first two pseudoscalar states remain massless, the third
one (the
$\eta '$--meson) acquires a mass. In this sense there is a correspondence
between the vector mesons and the neutrinos, while the pseudoscalar mesons
correspond to the charged leptons. In particular the $\eta '$--meson
corresponds to the $\tau $--lepton.\\

Of course, this correspondence is simply given by the similarity of the mass
matrices and the underlying symmetry. The dynamics in the two situations is
quite different. In the case of the mesons the QCD dynamics reproduces the
mass and mixing pattern discussed above. In the case of leptons we can only
speculate, but the strong mass hierarchy of the charged leptons, which is at
least qualitatively similar to the mass hierarchy in the pseudoscalar channel,
may serve as a guide.\\

In any case the mixing pattern in the leptonic channel is of high interest for
neutrino oscillations. In our scheme the electron neutrino is in the limit
of $S(3)$--symmetry a superposition
of two mass eigenstates:
\begin{equation}
\nu_e = \frac{1}{\sqrt{2}} \left( \nu_1 - \nu_2 \right) \, .
\end{equation}

The mixing angle $\Theta $ is 45$^{\circ }$, i. e.
${\rm sin}^2 2 \Theta = 1$.
In terms of mass eigenstates the $\mu$-- and $\tau$--neutrinos are
given by:
\begin{eqnarray}
\nu_{\mu} & = & \frac{1}{\sqrt{6}} \left( \nu_1 + \nu_2 - 2 \nu_3 \right)
\nonumber \\
\nu_{\tau} & = & \frac{1}{\sqrt{3}} \left( \nu_1 + \nu_2 + \nu_3 \right) \, .
\end{eqnarray}

A $\mu$--neutrino will in general oscillate into all three neutrino mass
eigenstates. However,
due to the high degeneracy between the $\nu_1$ and $\nu_2$--states,
oscillations between $\mu$--neutrinos and electron neutrinos will appear
only at very large distances. Oscillations between $\mu$--neutrinos and
$\tau$--neutrinos could show up at
smaller distances, if the mass difference between the $\left( \nu_1, \nu_2
\right)$--states and the $\nu_3$--state is sizeable. For the sake of our
discussion let us suppose that the first two neutrino states are completely
degenerate, in which case we can perform a 45$^{\circ}$--rotation among the
two states without changing the physical situation.\\

Denoting the state $\left( \nu_1 - \nu_2 \right) / \sqrt{2}$ by $\tilde{\nu}_1$
and $\left( \nu_1 + \nu_2 \right) / \sqrt{2}$ by $\tilde{\nu}_2$, one finds
the doublets:
\begin{equation}
\left( {\tilde{\nu}_1 \atop e^-} \right) \qquad \left(
{\frac{1}{\sqrt{3}} \tilde{\nu}_2 - \sqrt{\frac{2}{3}} \nu_3 \atop \mu^-}
\right) \qquad \left(
{\sqrt{\frac{1}{3}} \tilde{\nu}_2 + \frac{1}{\sqrt{3}} \nu_3 \atop \tau^-}
\right)
\end{equation}

Effectively the mixing angle between the $\mu - \tau$ system is given by
${\rm arc} {\rm sin} \sqrt{\frac{2}{3}} = 54.7^{\circ}$, which gives
${\rm sin}^2 2 \Theta = \frac{8}{9}$.\\

The atmospheric neutrino experiments are consistent with a large mixing angle
describing $\nu_{\mu} \leftrightarrow \nu_{\tau}$ oscillations:
\begin{equation}
P \left( \nu_{\mu} \rightarrow \nu_{\tau} \right) = {\rm sin}^2 \, 2
\Theta_{{\rm atm}} {\rm sin}^2 \left( 1.27
\frac{\Delta m^2_{{\rm atm}}L}{|P|} \right)
\end{equation}

with ${\rm sin}^2 2 \Theta_{{\rm atm}} \approx 1$ and $\Delta
m^2_{{\rm atm}} \approx 10^{-3} \, \, \, {\rm eV} \, \, \, ^2$. In our
case we
obtain ${\rm sin}^2 2 \Theta_{{\rm atm}} = \frac{8}{9}$ in the absence of
violation of the underlying $S(3)$ symmetry.\\
\\
The the observational hints towards neutrino oscillations both for solar and
atmospheric neutrinos indicate a mass pattern for the three neutrino states
as follows. The first two neutrinos $\nu_1$ and $\nu_2$ are nearly degenerate,
while the mass of the third neutrino $\nu_3$ is larger or smaller.\\
\\
Violations of the underlying $S(3)$--symmetry due to the non--zero masses for
the muon and the electron lead to small modifications of the mixing angles
given above. These departures from the symmetry limit depend on details
of the symmetry breaking and will not be discussed here$^{15}$. However,
the gross features of the observed oscillation pattern point towards
large mixing angles very similar to the large mixing angles seen in the
case of the $0^{-+}$--mesons:
\begin{eqnarray}
{\rm tan} \, ^2 \Theta ({\rm sun}) & \approx & 0.30 - 0.58 \nonumber \\
\nonumber \\
{\rm sin} \, ^2 2 \Theta {\rm (atm)} & > & 0.92.
\end{eqnarray}
\\
Both the experimental data as well as our theoretical considerations suggest
that in the case of leptons large mixing angles appear, just as in the case
of the pseudoscalar mesons in comparison to the vector mesons. The vector
mesons are nearly ``pure'' in terms of quark states, while the pseudoscalar
mesons are strongly mixed. Likewise in our approach the neutrinos are
``pure'' in terms of $S(3)$--symmetry eigenstates, while the charged leptons
are strongly mixed. The mismatch between the two sectors is the physical
origin of neutrino oscillations.
The mass and mixing pattern of the mesons is well understood within the
theory of QCD. It remains to be seen whether the $S(3)$--symmetry
eigenstates of quarks and leptons used in our analysis reflect in an
analogous way that these states
are more fundamental than the mass eigenstates. Obviously an answer to this
question can only be given in a theory which goes beyond the boundaries of the
Standard Model. In the case of the mesons the mixing pattern gave important
clues towards a more fundamental picture of hadronic physics based on quarks
and QCD. Likewise a deeper understanding of the flavor mixing of the leptons
and quarks might provide the first view beyond the frontiers of the
Standard Model.

\vspace*{1cm}

\begin{center}\underline{References}\end{center}

\begin{enumerate}

\item M. Gell--Mann, Phys. Rev. 125 (1962) 1067.

\item Y. Ne\a'{e}man, Nucl. Phys. 26 (1961) 222.

\item See e. g.: S. Gasiorowicz, Elementary Particle Physics,
New York (1966).

\item M. Gell--Mann, Phys. Lett. 8 (1964) 214--215.

\item G. Zweig, CERN Reports 8182/Th. 401 and 8419/TH. 412,
January 17 and February 21 (1964); Reprinted in Developments in the Quark
Theory of Hadrons, A Reprint Collection, Vol. I: 1964--1978,
Don.~B.~Lichtenberg and S. Peter Rosen editors, Hadronic Press, Nonantum
Massachusetts, 1980.

\item G. Zweig, private communication (1999).

\item P. L. Connolly et al., Phys. Rev. Lett. 10 (1963) 371.

\item Symmetries in Elementary Particle Physics, Academic Press,
New York \& London (1965), A.~Zichichi ed.

\item H. Fritzsch and P. Minkowski, Nuov. Cim. 30 (1975) 393.

\item H. Fritzsch, M. Gell--Mann and H. Leutwyler,
Phys. Lett. 47B (1973) 365.

\item G. 't Hooft, Phys. Rev. D14 (1976) 3432.

\item R. P. Feynman, private communication.

\item H. Fritzsch, Phys. Lett. B184 (1987) 391.

\item The first proposal for a third charged lepton with its own neutrino
is due
to A. Zichichi (see ``The Origin of the Third Family", World Scientific
Series in
the 20th Century Physics, Vol. 20, 1998. O. Barnabei, L. Maiani, R.A. Ricci
and F.
Roversi Monaco, eds.).

\item H. Fritzsch and Z. Z. Xing, Phys. Lett. B372 (1996) 265;\\
H. Fritzsch and Z. Z. Xing, Phys. Lett. B440 (1998) 313.
\end{enumerate}
\end{document}